\documentclass{ws-mpla}
\usepackage{amsfonts,amssymb,eucal}
\usepackage{amsmath}
\usepackage{amscd}
\usepackage{latexsym} 
\numberwithin{equation}{section}
\usepackage{amssymb,pstricks}
\begin{document}

\newcommand{\bX}{{\partial X}}
\newcommand{\cC}{\mathcal{C}}
\newcommand{\cDzu}{{{\mathcal D}_\Phi^{0,1}(X)}}
\newcommand{\cN}{{\mathcal N}}
\newcommand{\cz}{{\mathbb C}}
\newcommand{\cun}{\cC^{\infty}}
\newcommand{\ind}{\mathrm{index}}
\newcommand{\nz}{{\mathbb N}}
\newcommand{\pc}{{\Psi_\Phi(\rz^4)}}
\newcommand{\psus}{{\Psi_{\mathrm{sus}({}^\Phi\!T^*S^2)-\phi}(S^3)}}
\newcommand{\produs}{\mathrm{prod}}
\newcommand{\ptsX}{{{}^\Phi\!T^*X}}
\newcommand{\ptX}{{{}^\Phi\!TX}}
\newcommand{\px}{{\partial_x}}
\newcommand{\rz}{{\mathbb R}}
\newcommand{\sgn}{\mathrm{sign}}
\newcommand{\Sf}{\mathrm{sf}}
\newcommand{\Spec}{\mathrm{Spec}}
\newcommand{\tf}{\tilde{f}}
\newcommand{\tn}{\tilde{\nabla}}
\newcommand{\Tr}{\operatorname{Tr}}
\newcommand{\zz}{{\mathbb Z}}
\newcommand{\vf}{\varphi}
\newcommand{\be}{\begin{equation}}
\newcommand{\ee}{\end{equation}}
\newcommand{\ba}{\begin{eqnarray}}
\newcommand{\ea}{\end{eqnarray}}
\newcommand{\no}{\nonumber\\}
\renewcommand{\cal}{\mathcal}
\newcommand{\bx}{{\mathbf x}}
\newcommand{\by}{\bm{y}}
\renewcommand{\hat}{\widehat}
\newcommand {\ve}{\varepsilon}
\newcommand {\pr}{\partial}
\newcommand {\cG}{\cal G}
\newcommand {\cD}{\cal D}
\newcommand {\cL}{\cal L}
\newcommand {\G}{\Gamma}
\newcommand {\bg}{\bar \gamma}
\newcommand {\bp}{\bar \psi}
\newcommand {\bv}{\bar v}
\newcommand {\p}{\psi}
\newcommand {\vt}{\vartheta}
\title{Bel-Robinson tensor and dominant energy property in \\ 
the Bianchi type I Universe}

\author{\footnotesize BIJAN SAHA\footnote{
E-mail:~~~ saha@thsun1.jinr.ru,
http://www.jinr.ru/\~\,bijan} ,
VICTOR RIKHVITSKY}

\address{Laboratory of Information Technologies\\ 
Joint Institute for Nuclear Research\\
141980 Dubna, Moscow region, Russia} 

\author{\footnotesize MIHAI VISINESCU
\footnote{ 
E-mail:~~~ mvisin@theory.nipne.ro}}

\address{ Department of Theoretical Physics\\
National Institute for Physics and Nuclear Engineering\\
Magurele, P.O.Box MG-6, RO-077125 Bucharest, Romania}

\date{ }
\maketitle

\begin{abstract}
Within the framework of Bianchi type-I space-time we study the
Bel-Robinson tensor and its impact on the evolution of the
Universe. We use different definitions of the Bel-Robinson tensor
existing in the literature and compare the results. Finally we
investigate the so called "dominant super-energy property" for the
Bel-Robinson tensor as a generalization of the usual dominant
energy condition for energy momentum tensors.


Pacs: 03.65.Pm and 04.20.Ha

\end{abstract}

\section{Introduction}

The lack of a well-posed definition of {\it local energy-momentum
tensor} is the consequence of the Principle of Equivalence
\cite{misnergrav}, which lies at the heart of Einstein's theory of
general relativity. Nevertheless, quest for the local tensors
describing the strength of gravitational field has long been going
on. One of the first successful attempt to address this problem
was taken by Bel \cite{Bel1,Bel2,Bel3} and independently Robinson
\cite{Robinson}. In their works, in analogy with the
electromagnetic energy-momentum tensor, they constructed a
four-index tensor for the gravitational field in vacuum. The
properties of the now famous Bel-Robinson (BR) tensor are similar
to the traditional energy-momentum tensor and following Senovilla
\cite{Sen1,Sen2} can be formulated as follows: (i) it possesses a
positive-definite time-like component and a "causal" momentum
vector; (ii) its divergence vanishes (in vacuum); (iii) the tensor
is zero if and only if the curvature of the space-time vanishes;
(iv) it has positivity property similar to the electromagnetic
one; and some others. Construction of BR and the study of its
properties were widely considered by a number of authors, e.g.,
Desher {\it et. al.} \cite{Deser,Deser1}, Teyssandier \cite{Tey},
Senovilla \cite{Sen1}, Bergqvist \cite{GB}, Andersson
\cite{ander}, Wingbrant \cite{wing}, Choquet-Bruhat {it et. al.}
\cite{cho} etc.

It should be noted the authors of the papers mentioned above
considered the BR and established its properties in general.
On the other hand, in general relativity there exists a number 
of interesting 
and widely studied models of space-time. Therefore, in our view
it is interesting to consider the BR within the scope of some
concrete metric. In a recent paper \cite{BRs} we studied the BR
within the framework of Bianchi type I (BI) universe using two
different definitions. The purpose of this paper is to extend
that study for some other definitions and analyze the dominant
energy property (DEP) and dominant super-energy property (DSEP) 
within this model.

\section{Bianchi I Universe: a brief description}

A Bianchi type-I (BI) universe is the straightforward
generalization of the flat Robertson-Walker (RW) universe and is
one of the simplest models of an anisotropic universe that
describes a homogeneous and spatially flat universe. It has the
agreeable property that near the singularity it behaves like a
Kasner universe, even in the presence of matter, and consequently
falls within the general analysis of the singularity given by
Belinskii et al.~\cite{belinskii}. Also in a universe filled with
matter for $p\,=\,\zeta\,\ve, \quad \zeta < 1$, it has been shown
that any initial anisotropy in a BI universe quickly dies away and
a BI universe eventually evolves into a Friedmann-RW (FRW)
universe~\cite{jacobs}. Since the present-day universe is
surprisingly isotropic, this feature of the BI universe makes it a
prime candidate for studying the possible effects of an anisotropy
in the early universe on present-day observations. In light of the
importance mentioned above, several authors have studied BI
universe from different aspects.

A diagonal BI space-time is a spatially
homogeneous space-time, which admits an Abelian group $G_3$,
acting on spacelike hypersurfaces, generated by the spacelike
Killing vectors $\bx_1 = \pr_{1},\, \bx_2 = \pr_{2}$, and $\bx_3 =
\pr_{3}$. In synchronous coordinates, the metric is
\cite{bianchi,tsam}:
\begin{equation}
ds^2 = dt^2 - \sum_{i=1}^{3}a_i^2 (t) dx_i^2. \label{BI}
\end{equation}

If the three scale factors are equal (i.e., $ a_1 = a_2 = a_3$),
Eq. (\ref{BI}) describes an isotropic and spatially flat
FRW universe. The BI universe has a
different scale factor in each direction, thereby introducing an
anisotropy to the system. Thus, a BI universe,
being the straightforward generalization of the flat FRW universe,
is one of the simplest models of an anisotropic universe that
describes a homogeneous and spatially flat universe. When two of
the metric functions are equal (e.g., $a_2 = a_3$) the BI
space-time is reduced to the important class of plane symmetric
space-time (a special class of the locally rotational symmetric
space-times~\cite{ellis1,ellis2}), which admits a $G_4$ group of
isometries acting multiply transitively on the spacelike
hypersurfaces of homogeneity generated by the Killing vectors
$\bx_1,\,\bx_2,\,\bx_3$, and $\bx_4 = x^2\pr_{3} - x^3\pr_{2}$.
The BI has the agreeable property that near the singularity it
behaves like a Kasner universe, given by
\begin{equation}
a_1(t) = a_{1}^{0} t^{p_1}, \quad a_2(t) = a_{2}^{0} t^{p_2},
\quad a_3(t) = a_{3}^{0} t^{p_3}, 
\nonumber
\end{equation}
with $p_j$ being the parameters of the BI space-time which measure
the relative anisotropy between any two asymmetry axes and satisfy
the constraints
\begin{eqnarray}
p_1 + p_2 + p_3 = 1, \quad p_{1}^{2} + p_{2}^{2} + p_{3}^{2} = 1.
\label{kc}
\end{eqnarray}

As one sees, $p_1$, $p_2$ and $p_3$ cannot be equal. Only two of
them can be equal, and only in two special cases, namely,
$(0,0,1)$ and $(-1/3,2/3,2/3)$. In all other cases $p_1$, $p_2$
and $p_3$ are different, moreover, one of them is negative, while
the two others are positive. If it is supposed that $p_1 < p_2 <
p_3$, then their values are confined in the following intervals:
$$ -1/3 \le p_1 \le 0, \quad 0 \le p_2 \le 2/3, \quad 2/3 \le p_3
\le 1.$$ The solutions of the algebraic equations (\ref{kc}) can
be presented as
\begin{eqnarray}
p_{1} = \frac{-p}{p^2 + p +1},\quad p_{2} = \frac{p(p + 1)}{p^2 +
p +1},\quad p_{3} = \frac{p+1}{p^2 + p +1}. 
\nonumber
\end{eqnarray}
Thus instead of three, we have now one parameter $p$, which lies
in the interval $0 \le p \le 1$.

Another particular parametrization can be given using an angle on
the unit circle, since Eqs. (\ref{kc}) describe the intersection
of a sphere with a plane in the parameter space $(p_1,p_2,p_3)$:
\begin{eqnarray}
p_{1} = \frac{1 + {\rm cos} \vt + \sqrt{3}{\rm sin} \vt}{3},\quad
p_{2} = \frac{1 + {\rm cos} \vt - \sqrt{3}{\rm sin} \vt}{3},\quad
p_{3} = \frac{1 - 2 {\rm cos} \vt}{3}. 
\nonumber
\end{eqnarray}
Although $\vt$ ranges over the unit circle, the labeling of each
$p_j$ is quite arbitrary. Thus the unit circle can be divided into
six equal parts, each of which span $60^{\circ}$, and the choice
of $p_j$ is unique within each section separately. For $\vt =0$,
$p_1 = p_2 = \frac{2}{3}$ and $p_3 = -\frac{1}{3}$ while for $\vt
= \pi/3,$ $p_1 = 1$ and $p_2 = p_3 = 0$.

For later convenience we list the Christoffel symbol, scalar 
curvature, Ricci, Riemann and Weyl tensors for the BI space-time. 
The non-trivial Christoffel symbols are the following:
\begin{equation}
\G^{i}_{i0} = \frac{\dot a_i}{a_i}, \quad \G^{0}_{ii} = a_i {\dot
a_i}, \quad i=1,2,3\, , 
\nonumber
\end{equation}
and the non-trivial components of Riemann tensors are
\begin{eqnarray}
R^{0i}_{\,\,\,\,\,\,\,\,0i} &=& - \frac{\ddot a_i}{a_i}, \quad
R^{ij}_{\,\,\,\,\,\,\,\,ij} = -\frac{\dot a_i}{a_i}\frac{\dot
a_j}{a_j}, \label{riemann} \\
& & i, j =1,2,3, \quad i \ne j.
\nonumber
\end{eqnarray}

Finally the nontrivial components of the Ricci tensors  for the BI metric
are
\begin{eqnarray}
R_{0}^{0} = - \sum_{i=1}^{3} \frac{\ddot a_i}{a_i},\quad R_{i}^{i}
&=& - \Bigl[\frac{\ddot a_i}{a_i} + \frac{\dot a_i}{a_i}\Bigl(
\frac{\dot a_j}{a_j} + \frac{\ddot a_k}{a_k}\Bigr)\Bigr],
\no
 & & i, j, k = 1,2,3, \quad i \ne j \ne k. \nonumber
\end{eqnarray}
and the scalar curvature is
\begin{equation}
R = - 2\Bigl(\sum_{i=1}^{3} \frac{\ddot a_i}{a_i}+ \frac{\dot
a_1}{a_1}\frac{\dot a_2}{a_2} + \frac{\dot a_2}{a_2}\frac{\dot
a_3}{a_3}+ \frac{\dot a_3}{a_3}\frac{\dot a_1}{a_1}\Bigr).
\nonumber
\end{equation}

It is convenient to separate the Riemann tensor into a trace-free
part and a "Ricci" part. This gives the Weyl tensor
\begin{eqnarray}
C_{ijkl} &=& R_{ijkl} - \frac{1}{(n-2)} \Bigl( g_{ik} R_{jl} +
g_{jl} R_{ik} - g_{jk} R_{il} - g_{il} R_{jk}\Bigr)   \no
&+& \frac{1}{(n-1)(n-2)} \Bigl(g_{ik}g_{jl} -g_{il}g_{jk}\Bigr) R.
\nonumber
\end{eqnarray}

This tensor has manifestly all the symmetries of the Riemann
tensor; but contrary to the Riemann tensor while it gives rise to
Ricci tensor, the Weyl tensor gives
\begin{equation}
g^{ik}C_{ijkl} \equiv 0. \label{Weytr} 
\end{equation}

A further distinction is that while the Riemann tensor can be
defined in a manifold endowed only with a connection, the Weyl
tensor can be defined only when a metric is also defined. In 4 
dimensions 
the Riemann tensor has 20 distinct components, while the Weyl and
the Ricci have 10 components each. The non-trivial components of
the Weyl tensor for the BI space-time are
\begin{eqnarray}
C_{0i0i} &=&\frac{a_i}{6 a_j a_k} \bigl\{2 {\ddot a_i} a_j a_k -
{\ddot a_j} a_k a_i - {\ddot a_k} a_i a_j - {\dot a_i} {\dot a_j}
a_k - {\dot a_k} {\dot a_i} a_j + 2 {\dot a_j} {\dot a_k} a_i
\bigr\} \no C_{jkjk} &=&-\frac{a_j a_k}{6 a_i} \bigl\{2 {\ddot
a_i} a_j a_k - {\ddot a_j} a_k a_i - {\ddot a_k} a_i a_j - {\dot
a_i} {\dot a_j} a_k - {\dot a_k} {\dot a_i} a_j + 2 {\dot a_j}
{\dot a_k} a_i \bigr\},\no & & i, j, k =1,2,3, \quad i \ne j \ne k
. \label{Weyl}
\end{eqnarray}
{}From \eqref{Weyl} one easily finds the following relation:
\begin{equation}
C_{0i0i} = - \frac{a_i^2}{a_j^2 a_k^2} C_{jkjk}, \quad i,j,k =
1,2,3, \quad i \ne j \ne k. \label{Weylrel}
\end{equation}

Now having all the non-trivial components of Ricci and Riemann
tensors, one can easily write the invariants of gravitational
field which we need to study the space-time singularity. Moreover
now we can construct the BR tensor that is defined
differently by different authors.

\section{Einstein equations and their solutions}

In this section we study the Einstein equation. In doing so let us
first write the Einstein equation for the BI metric governing the
evolution of the Universe. In presence of a cosmological constant
$\Lambda$ the Einstein equation has the form
\begin{subequations}
\begin{eqnarray}
\frac{\ddot a_2}{a_2} +\frac{\ddot a_3}{a_3} + \frac{\dot
a_2}{a_2}\frac{\dot a_3}{a_3}&=&  \kappa T_{1}^{1} +
\Lambda,\label{11}\\ \frac{\ddot a_3}{a_3} +\frac{\ddot a_1}{a_1}
+ \frac{\dot a_3}{a_3}\frac{\dot a_1}{a_1}&=& \kappa T_{2}^{2}
+\Lambda,\label{22}\\ \frac{\ddot a_1}{a_1} +\frac{\ddot a_2}{a_2}
+ \frac{\dot a_1}{a_1}\frac{\dot a_2}{a_2}&=&  \kappa T_{3}^{3} +
\Lambda,\label{33}\\ \frac{\dot a_1}{a_1}\frac{\dot a_2}{a_2}
+\frac{\dot a_2}{a_2}\frac{\dot a_3}{a_3} +\frac{\dot
a_3}{a_3}\frac{\dot a_1}{a_1}&=&  \kappa T_{0}^{0} + \Lambda.
\label{00}
\end{eqnarray}
\end{subequations}
Here over-dot means differentiation with respect to $t$ and
$T_\mu^\nu$ is the energy-momentum tensor of the matter field
which we choose in the form:
\begin{equation}
T_{\mu}^{\nu} = (\ve + p) u_\mu u^\nu - p \delta_\mu^\nu,
\label{emt}
\end{equation}
where $u^\mu$ is the flow vector satisfying
\begin{equation}
g_{\mu\nu} u^\mu u^\nu = 1. 
\nonumber
\end{equation}
Here $\ve$ is the total energy density of a perfect fluid and/or
dark energy density, while $p$ is the corresponding pressure. $p$
and $\ve$ are related by an equation of state which will be
studied below in detail. In a co-moving system of coordinates from
(\ref{emt}) one finds
\begin{equation}
T_0^0 = \ve, \qquad T_1^1 = T_2^2 = T_3^3 = - p. \label{compemt}
\end{equation}

In view of (\ref{compemt}) from (\ref{11}) - (\ref{00}) one
immediately obtains \cite{PRD23501}
\begin{eqnarray}
a_i(t) &=& D_{i} [\tau(t)]^{1/3} \exp \bigl[X_i \int\,
[\tau(t')]^{-1}  dt' \bigr], \quad i=1,2,3. \nonumber
\end{eqnarray}
Here $D_i$ and $X_i$ are some arbitrary constants obeying $$D_1
D_2 D_3 = 1, \qquad X_1 + X_2 + X_3 = 0,$$ and $\tau$ is a
function of $t$ defined to be
\begin{equation}
\tau = a_1 a_2 a_3. 
\nonumber
\end{equation}
{}From (\ref{11}) - (\ref{00}) for $\tau$ one find
\begin{equation}
\frac{\ddot \tau}{\tau} = \frac{3 \kappa}{2} \bigl(\ve - p\bigr) +
3 \Lambda. \label{dtau}
\end{equation}
On the other hand the conservation law for the energy-momentum
tensor gives
\begin{equation}
\dot{\ve} = -\frac{\dot \tau}{\tau} \bigl(\ve + p\bigr).
\label{dve}
\end{equation}
After a little manipulations from (\ref{dtau}) and (\ref{dve}) we
find
\begin{equation}
\dot{\tau}^2 = 3 (\kappa\ve + \Lambda) \tau^2 + C_1, \label{tve}
\end{equation}
with $C_1$ being an arbitrary constant. Let us now, in analogy
with Hubble constant, define
\begin{equation}
\frac{\dot \tau}{\tau} =  \frac{\dot a}{a} + \frac{\dot b}{b} +
\frac{\dot c}{c} = 3 H. \label{Hubble}
\end{equation}
On account of (\ref{Hubble}) from (\ref{tve}) one derives
\begin{equation}
\kappa \ve = 3 H^2 - \Lambda - C_1/(3 \tau^2). \label{tven}
\end{equation}

It should be noted that the energy density of the Universe is a
positive quantity. It is believed that at the early stage of
evolution when the volume scale $\tau$ was close to zero, the
energy density of the Universe was infinitely large. On the other
hand with the expansion of the Universe, i.e., with the increase
of $\tau$, the energy density $\ve$ decreases and an infinitely
large $\tau$ corresponds to a $\ve$ close to zero. Say at some
stage of evolution $\ve$ is too small to be ignored. In that case
from (\ref{tven}) follows
\begin{equation}
3 H^2 - \Lambda \to 0.\label{limit}
\end{equation}

As it is seen  from (\ref{limit}) in this case $\Lambda$ is
essentially non-negative. We can also conclude from (\ref{limit})
that in absence of a $\Lambda$ term beginning from some value of
$\tau$ the evolution of the Universe comes stand-still, i.e.,
$\tau$ becomes constant, since $H$ becomes trivial, whereas in
case of a positive $\Lambda$ the process of evolution of the
Universe never comes to a halt. Moreover it is believed that the
presence of the dark energy (which can be explained with a
positive $\Lambda$ as well) results in the accelerated expansion
of the Universe. As far as negative $\Lambda$ is concerned, its
presence imposes some restriction on $\ve$, namely, $\ve$ can
never be small enough to be ignored. In case of the perfect fluid
given by $p = \zeta \ve$ there exists some upper limit for $\tau$
as well (note that $\tau$ is essentially nonnegative, i.e. bound
from below). In our previous papers we came to the same conclusion
\cite{PRD23501,PRD24010} (with a positive $\Lambda$ which in the
present paper appears to be negative). A suitable choice of
parameters in this case may give rise to an oscillatory mode of
expansion, whereas in case of a Van der Waals fluid the highly
nonlinear equation of state may result in an exponential expansion
as well.

Inserting (\ref{Hubble}) and (\ref{tven}) into (\ref{dtau}) one
now finds
\begin{equation}
\dot H = - \frac{1}{2}\bigl(3 H^2 - \Lambda + \frac{C_1}{3 \tau^2}
+ \kappa p\bigr) = -\frac{\kappa}{2} \bigl(\ve + p\bigr) -
\frac{C_1}{3 \tau^2}\,. \label{Hdef}
\end{equation}
In view of (\ref{tven}), from (\ref{Hdef}), it follows that if the
perfect fluid is given by a stiff matter where $p = \ve$, the
corresponding solution does not depend on the constant $C_1$.

Let us now go back to the Eq. (\ref{tve}). It is in fact the first
integral of (\ref{dtau}) and can be written as
\begin{equation}
\dot \tau = \pm \sqrt{C_1 + 3(\kappa \ve + \Lambda) \tau^2}
\label{fi}
\end{equation}

On the other hand, rewriting (\ref{dve}) in the form
\begin{equation}
\frac{\dot\ve}{\ve + p} = \frac{\dot \tau}{\tau}, \nonumber
\end{equation}
and taking into account that $p$ is a function of $\ve$, one
concludes that the right hand side of the Eq. (\ref{dtau}) is a
function of $\tau$ only, i.e.,
\begin{equation}
\ddot \tau = \frac{3 \kappa}{2} \bigl(\ve - p\bigr) \tau + 3
\Lambda \tau = \mathcal F(\tau). \label{ddtau}
\end{equation}

{}From a mechanical point of view Eq. (\ref{ddtau}) can be
interpreted as an equation of motion of a single particle with
unit mass under the force $\mathcal F(\tau)$. Then the following
first integral exists \cite{PRD24010}:
\begin{equation}
    \dot \tau = \sqrt{2[\mathcal E - \mathcal U(\tau)]}\,.
\label{1stint}
\end{equation}
Here $\mathcal E$ can be viewed as energy and $\mathcal U(\tau)$
is the potential of the force $\mathcal F$. Comparing the Eqs.
(\ref{fi}) and (\ref{1stint}) one finds $\mathcal E = C_1/2$ and
\begin{equation}
\mathcal U(\tau) = -\frac{3}{2}(\kappa \ve + \Lambda) \tau^2.
\label{poten}
\end{equation}

Let us finally write the solution to the Eq. (\ref{dtau}) in
quadrature:
\begin{equation}
\int \frac{d\tau}{\sqrt{C_1 + 3(\kappa \ve + \Lambda) \tau^2}} = t
+ t_0, \label{quad}
\end{equation}
where the integration constant $t_0$ can be taken to be zero,
since it only gives a shift in time. The Eqs. (\ref{dtau}) and
(\ref{dve}) for perfect fluid obeying different equations of state
has been thoroughly studied by us \cite{PRD23501,PRD24010}.

\section{Bel-Robinson tensors}

BR tensor first appeared in the endless search for
a covariant version of gravitational energy. In general
relativity, the energetic content of an electromagnetic field
propagating in a region free of charge is described by the
well-known symmetric trace-less tensor
\begin{equation}
T^{\alpha \beta}_{\rm el} = - \frac{1}{4\pi} \bigl(F^{\alpha
\lambda} F^{\beta}_{\lambda} - \frac{1}{4} g^{\alpha\beta}
F^{\mu\nu}F_{\mu\nu}\bigr),\label{emtel}
\end{equation}
where $F^{\alpha \beta}$ is the electromagnetic field  tensor.
 This tensor satisfies:
\begin{equation}
T^{\alpha \beta}_{\rm el\,;\alpha} = 0 
\nonumber
\end{equation}
as a consequence of Maxwell equations with $j^{\mu} = 0.$ The
tensor $T^{\alpha \beta}_{\rm el}$ enables us to define a local
density of electromagnetic energy as measured by an observer
moving with the unit 4-velocity $u$:
\begin{equation}
w_{\rm el} (u) = T^{\alpha \beta}_{\rm el} u_\alpha u_\beta.
\nonumber
\end{equation}
It follows from (\ref{emtel}) that the energy density is positive
definite for any time-like vector $u$.

Within the scope of general relativity, however, it is well known
that the concept of local energy density is meaningless for a
gravitational field. To overcome this difficulty led to introduce
the notion of super-energy tensor constructed with the curvature
tensor $R_{\mu\nu\alpha\beta}$. The first example of such a tensor
was exhibited by Bel \cite{Bel1}, that was further generalized to
the case of an arbitrary gravitational field \cite{Bel2}. Note
that a similar tensor was also introduced by Robinson
\cite{Robinson}. This tensor is now commonly know as the
BR tensor as well. Since we are going to compare
some distinct definition of BR in this paper, before defining them
let us see what kind of properties they should have. In general,
the BR tensor has the following symmetry properties:
\begin{subequations}
\begin{eqnarray}
B_{\mu\nu\alpha\beta} &=& B_{\nu\mu\alpha\beta}, \label{1pair}
\\ B_{\mu\nu\alpha\beta} &=& B_{\mu\nu\beta\alpha},\label{2pair}
\\ B_{\mu\nu\alpha\beta} &=& B_{\alpha\beta\mu\nu}.\label{both}
\end{eqnarray}
\end{subequations}
The symmetry property leads to the fact that that in
$n$-dimensional case there are $n(n+1)[n(n+1)+2]/8$ independent
components of the BR tensor. In case of $n = 4$ out of
256 components only 55 are linearly independent.

In literature there are a few definitions of BR. Here we mention
only three.

{\bf I.} By analogy with the tensor (\ref{emtel}) which may be
written as
\begin{equation}
T_{\mu \nu} = F_{\mu \alpha} F_{\nu}^{\alpha} + \ast F_{\mu
\alpha} \ast F_{\nu}^{\alpha}, 
\nonumber
\end{equation}
the BR tensor is defined as \cite{Deser}:
\begin{equation}
B_{\mu\nu\alpha\beta} = R^{\rho \,\,\, \sigma}_{\,\,\,\mu
\,\,\,\, \alpha} R_{\rho \nu \sigma \beta} + \ast R^{\rho \,\,\,
\sigma} _{\,\,\,\mu \,\,\,\, \alpha} \ast R_{\rho \nu \sigma
\beta}. \label{Bel1}
\end{equation}
Here the dual curvature is $\ast R^{\mu
\nu}_{\,\,\,\,\,\,\,\,\,\,\lambda \sigma} \equiv (1/2)\,
\epsilon^{\mu \nu}_{\,\,\,\,\,\,\,\,\,\alpha \beta} R^{\alpha
\beta}_{\,\,\,\,\,\,\,\,\,\lambda \sigma}.$ It should be noted
that this definition is adequate only in 4 dimensions and in 
vacuum. Otherwise this tensor cannot satisfy the DEP \cite{sen-pc} 
and therefore this expression should not be used in other 
dimensions or in non-Ricci-flat spacetimes.

Using the definition of dual curvature, from (\ref{Bel1}) we find
\begin{equation}
B_{\mu\nu\alpha\beta} = R^{\rho\,\,\, \sigma}_{\,\,\,\mu
\,\,\,\,\alpha} R_{\rho \nu \sigma \beta} + R^{\rho \,\,\,
\sigma}_{\,\,\,\mu \,\,\,\,\beta} R_{\rho \nu \sigma \alpha} -
\frac{1}{2} g_{\mu \nu} R_{\alpha}^{\,\,\,\,\,\rho\sigma\tau}
R_{\beta\rho\sigma \tau}. \label{Bel2}
\end{equation}

The properties (\ref{1pair}) and (\ref{2pair}) follow immediately
from (\ref{Bel1}) thanks to the symmetry property of Riemann
tensor. The property (\ref{both}) is straightforward from
(\ref{Bel1}), but for (\ref{Bel2}) it requires
\begin{equation}
g_{\mu \nu} R_{\alpha}^{\,\,\,\,\,\rho\sigma\tau}
R_{\beta\rho\sigma \tau} = g_{\alpha \beta}
R_{\mu}^{\,\,\,\,\,\rho\sigma\tau} R_{\nu\rho\sigma \tau}.
\label{restric}
\end{equation}
In view of (\ref{riemann}) for the BR tensor in this
case we obtain the following non-trivial components:
\begin{eqnarray}
B_{0000} &=& \sum_{i=1}^{3} \frac{{\ddot a_i}^2}{a_i^2},
\no 
B_{iiii} &=&{\ddot a_i}^2 + {\dot a_i}^2 \biggl\{
\frac{\dot a_j^2}{a_j^2} + \frac{\dot a_k^2}{a_k^2}\biggr\},
\no 
B_{0i0i} &=& a_i{\ddot a_i}\biggl\{\frac{\ddot
a_j}{a_j} + \frac{\ddot a_k}{a_k} \biggr\},
\no
B_{ijij} &=& {\ddot a_i}{\ddot a_j} + a_i {\dot a_i} a_j {\dot
a_j} \frac{{\dot a_k}^2}{a_k^2}, 
\no 
B_{00ii} &=&
{\ddot a_i}^2 - {\dot a_i}^2 \biggl\{\frac{{\dot a_j}^2}{a_j^2} +
\frac{{\dot a_k}^2}{a_k^2}\biggr\} , 
\no 
B_{iijj}&=&
{\dot a_i}^2 {\dot a_j}^2  -  a_i^2 {\ddot a_j}^2 - a_i^2 {\dot
a_j}^2\frac{{\dot a_k}^2}{a_k^2}, 
\no
 & & i, j, k = 1,
2, 3 \quad i\ne j \ne k. \nonumber
\end{eqnarray}

Inserting (\ref{riemann}) into (\ref{restric}) we obtain following
additional relations:
\begin{equation}
\biggl(\frac{\ddot a_i}{a_i}\biggr)^2 \pm \biggl(\frac{\ddot
a_j}{a_j}\biggr)^2 = \biggl(\frac{\dot a_k}{a_k}\biggr)^2 \biggl[
\biggl(\frac{\dot a_i}{a_i}\biggr)^2 \pm \biggl(\frac{\dot
a_j}{a_j}\biggr)^2\biggr], \quad i, j, k = 1, 2, 3 \quad i\ne j
\ne k \label{restric1}
\end{equation}
Among the six constrains in (\ref{restric1}) only three are
linearly independent. After a little manipulations with them
finally obtains the following relations between the metric
functions:
\begin{equation}
\frac{\ddot a_i}{a_i} = \frac{\dot a_j}{a_j}\frac{\dot
a_k}{a_k},\quad i, j, k = 1, 2, 3 \quad i\ne j \ne k.
\label{abcrel}
\end{equation}
As one sees, in account of (\ref{abcrel}) the Einstein equation
(\ref{11}) - (\ref{00}) leads to $T_0^0 = T_1^1 = T_2^2 = T_3^3$,
which can be realized only when the source field satisfies the
following equation of state:
\begin{equation}
p = -\ve. \label{eos}
\end{equation}

It is well known that only vacuum satisfies the state of equation
given by (\ref{eos}). Thus we see that if we are to define BR
tensor given by (\ref{Bel1}) or (\ref{Bel2}) we should deal with
the Einstein equations with the source field given by a vacuum.

{\bf II.} The restriction that arises above is due to the fact
that in defining the BR tensor we used the dual term
with the duality operator acting on the left pair only. To avoid
this restrictions the BR tensor can be defined by
\cite{Tey,Che}
\begin{eqnarray}
2 B_{\mu\nu\alpha\beta} &=& R^{\rho \,\,\, \sigma}_{\,\,\,\mu
\,\,\,\,\, \alpha} R_{\rho \nu \sigma \beta} + \ast R^{\rho \,\,\,
\sigma} _{\,\,\,\mu \,\,\,\,\, \alpha} \ast R_{\rho \nu \sigma
\beta} \no &+&  R \ast^{\rho \,\,\, \sigma} _{\,\,\,\mu \,\,\,\,\,
\alpha}  R \ast_{\rho \nu \sigma \beta} + \ast R \ast^{\rho \,\,\,
\sigma} _{\,\,\,\mu \,\,\,\,\, \alpha} \ast R \ast_{\rho \nu
\sigma \beta}, \label{Bel01}
\end{eqnarray}
where the duality operator acts on the left or on the right pair
of indices according to its position. Nowadays this is known as
the Bel tensor and was introduced by Bel \cite{Bel2} in a slightly
different form.

{}From (\ref{Bel01}) one easily finds
\begin{eqnarray}
B_{\mu\nu\alpha\beta} &=& R^{\rho\,\,\, \sigma}_{\,\,\,\mu
\,\,\,\,\,\alpha} R_{\rho \nu \sigma \beta} + R^{\rho \,\,\,
\sigma}_{\,\,\,\mu \,\,\,\,\,\beta} R_{\rho \nu \sigma \alpha} -
\frac{1}{2} g_{\mu \nu} R_{\alpha}^{\,\,\,\,\,\rho\sigma\tau}
R_{\beta\rho\sigma \tau} \no 
& - & \frac{1}{2} g_{\alpha
\beta} R_{\mu}^{\,\,\,\,\,\rho\sigma\tau} R_{\nu\rho\sigma \tau} +
\frac{1}{8} g_{\mu \nu} g_{\alpha \beta} R^{\rho\sigma\tau\eta}
R_{\rho\sigma \tau\eta}. 
\nonumber
\end{eqnarray}
Under the new definition the symmetry properties
(\ref{1pair}), (\ref{2pair}) and (\ref{both}) follow immediately,
without any restriction to the metric functions.

Let us now write the non-trivial components of the BR
tensor for the BI metric. In view of (\ref{riemann}) we now find
\begin{eqnarray}
B_{0000} &=& \frac{1}{2}\biggl\{\sum_{i=1}^{3} \frac{{\ddot
a_i}^2}{a_i^2} +\frac{{\dot a_1}^2}{a_1^2}\frac{{\dot
a_2}^2}{a_2^2} + \frac{{\dot a_2}^2}{a_2^2}\frac{{\dot
a_3}^2}{a_3^2} + \frac{{\dot a_3}^2}{a_3^2}\frac{{\dot
a_1}^2}{a_1^2}\biggr\},\no 
B_{iiii} &=& a_i^4 B_{0000},\no 
B_{0i0i}
&=& -a_i{\dot a_i}\biggl\{\frac{\dot a_j}{a_j}\frac{\ddot
a_j}{a_j} + \frac{\dot a_k}{a_k} \frac{\ddot a_k}{a_k}\biggr\},\no
B_{ijij} &=& a_i a_j \biggl\{{\ddot a_i}{\ddot a_j} +  {\dot
a_i}{\dot a_j} \frac{{\dot a_k}^2}{a_k^2}\biggr\},  \no
B_{00ii}
&=& \frac{1}{2}\biggl\{ -{\ddot a_i}^2 + {\dot a_i}^2
\biggl(\frac{{\dot a_j}^2}{a_j^2} + \frac{{\dot
a_k}^2}{a_k^2}\biggr) + a_i^2 \biggl(\frac{{\ddot a_j}^2}{a_j^2} +
\frac{{\ddot a_k}^2}{a_k^2}- \frac{{\dot a_j}^2}{a_j^2}\frac{{\dot
a_k}^2}{a_k^2}\biggr)\biggr\} , \no 
B_{iijj}&=&
\frac{1}{2}\biggl\{{\dot a_i}^2 {\dot a_j}^2  - {\ddot a_j}^2
a_j^2 - a_i^2 {\ddot a_j}^2 -\frac{{\dot a_k}^2}{a_k^2}
\bigl({\dot a_i}^2 a_j^2 + a_i^2 {\dot a_j}^2 \bigr) + a_i^2 a_j^2
\frac{{\dot a_k}^2}{a_k^2} \biggr\}, \no
 & & i, j, k = 1,
2, 3 \quad i\ne j \ne k. \nonumber
\end{eqnarray}
But the BR tensor defined in this way is not trace-free and is not
completely symmetric. It is achieved if and only if the manifold
is Ricci flat, i.e., $R_{ij} = 0$. Since for the BI universe we
have non-trivial components of Ricci tensor, we give an
alternative definition of BR where it is totally symmetric and
trace-free.

{\bf III.} Here we give another definition that gives rise to BR
tensor, that is trace-less and totally symmetric. It can be
achieved by constructing BR by means of Weyl tensor \cite{GB,Sen}.

\begin{equation}
B_{\mu\nu\alpha\beta} = C^{\rho \,\,\, \sigma}_{\,\,\,\mu
\,\,\,\,\alpha} C_{\rho \nu \sigma \beta} + \ast C^{\rho \,\,\,
\sigma} _{\,\,\,\mu \,\,\,\,\alpha} \ast C_{\rho \nu \sigma
\beta}. \label{BelW1}
\end{equation}
It can be shown that this BR is totally symmetric, i.e.,
\begin{equation}
B_{ijkl} = B_{(ijkl)}, 
\nonumber
\end{equation}
Moreover, as one can easily find from (\ref{Weytr}), the BR
defined through Weyl tensor is trace-free, i.e.,
\begin{equation}
g^{jl} B_{ijkl} \equiv 0. \label{BRtr}
\end{equation}

Let us study this case in detail. Using the properties of
Levi-Civita tensor we first rewrite \eqref{BelW1} in the form
\begin{equation}
B_{\mu\nu\alpha\beta} = C^{\rho\,\,\, \sigma}_{\,\,\,\mu
\,\,\,\,\,\alpha} C_{\rho \nu \sigma \beta} + C^{\rho \,\,\,
\sigma}_{\,\,\,\mu \,\,\,\,\,\beta} C_{\rho \nu \sigma \alpha} -
\frac{1}{2} g_{\mu \nu} C_{\alpha}^{\,\,\,\,\,\rho\sigma\tau}
C_{\beta\rho\sigma \tau}. 
\nonumber
\end{equation}
After a little manipulation one easily finds the following
nontrivial components of the BR tensor.
\begin{equation}
B_{0000} = \Bigl(g^{11} C_{1010}\Bigr)^2 + \Bigl(g^{22}
C_{2020}\Bigr)^2 +\Bigl(g^{33} C_{3030}\Bigr)^2, 
\nonumber
\end{equation}
which on account of \eqref{Weylrel} gives
\begin{equation}
B_{0000} = \Bigl(\frac{1}{a_2^2 a_3^2} C_{2323}\Bigr)^2 +
\Bigl(\frac{1}{a_3^2 a_1^2} C_{3131}\Bigr)^2 +
\Bigl(\frac{1}{a_1^2 a_2^2} C_{1212}\Bigr)^2. \label{Ab0f}
\end{equation}
On the other hand for we find
\begin{equation}
B_{1111} = \Bigl(C_{1010}\Bigr)^2 + \Bigl(g^{22} C_{1212}\Bigr)^2
+\Bigl(g^{33} C_{1313}\Bigr)^2, 
\nonumber
\end{equation}
which in view of \eqref{Weylrel} can be rewritten as
\begin{eqnarray}
B_{1111} &=& \Bigl(\frac{a_1^2}{a_2^2 a_3^2} C_{2323}\Bigr)^2 +
\Bigl(\frac{1}{a_3^2} C_{3131}\Bigr)^2
+ \Bigl(\frac{1}{a_2^2} C_{1212}\Bigr)^2 \label{Ab1f} \\
&=& a_1^4 \Biggl[\Bigl(\frac{1}{a_2^2 a_3^2} C_{2323}\Bigr)^2 +
\Bigl(\frac{1}{a_3^2 a_1^2} C_{3131}\Bigr)^2 +
\Bigl(\frac{1}{a_1^2 a_2^2} C_{1212}\Bigr)^2\Biggr] = a_1^4
B_{0000}.\nonumber
\end{eqnarray}
In view of \eqref{Ab0f} and \eqref{Ab1f} symbolically we can write
\begin{subequations}
\begin{eqnarray}
B_{0000} &=& \Bigl(\frac{1}{a_j^2 a_k^2} C_{jkjk}\Bigr)^2 +
\Bigl(\frac{1}{a_k^2 a_i^2} C_{kiki}\Bigr)^2 +
\Bigl(\frac{1}{a_i^2 a_j^2} C_{ijij}\Bigr)^2 \label{Ab0f0} \\
&=& \frac{1}{6}\biggl\{\sum_{i=1}^{3} \frac{{\ddot a_i}^2}{a_i^2}
+\frac{{\dot a_1}^2}{a_1^2}\frac{{\dot a_2}^2}{a_2^2} +
\frac{{\dot a_2}^2}{a_2^2}\frac{{\dot a_3}^2}{a_3^2} + \frac{{\dot
a_3}^2}{a_3^2}\frac{{\dot a_1}^2}{a_1^2} \no &-&\Bigl(\frac{{\ddot
a_1}}{a_1} \frac{{\ddot a_2}}{a_2} + \frac{{\ddot
a_2}}{a_2}\frac{{\ddot a_3}}{a_3} + \frac{{\ddot
a_3}}{a_3}\frac{{\ddot a_1}}{a_1}\Bigr) - \frac{{\dot
a_1}}{a_1}\frac{{\ddot a_2}}{a_2} \Bigl( \frac{{\ddot a_1}}{a_1} +
\frac{{\ddot a_2}}{a_2} - 2 \frac{{\ddot a_3}}{a_3} + \frac{{\dot
a_3}^2}{a_3^2}\Bigl) \no &-& \frac{{\dot a_2}}{a_2}\frac{{\ddot
a_3}}{a_3} \Bigl( \frac{{\ddot a_2}}{a_2} + \frac{{\ddot
a_3}}{a_3} - 2 \frac{{\ddot a_1}}{a_1} + \frac{{\dot
a_1}^2}{a_3^1}\Bigl) - \frac{{\dot a_3}}{a_3}\frac{{\ddot
a_1}}{a_1} \Bigl( \frac{{\ddot a_3}}{a_3} + \frac{{\ddot
a_1}}{a_1} - 2 \frac{{\ddot a_2}}{a_2} + \frac{{\dot
a_2}^2}{a_2^2}\Bigl) \biggr\}, \nonumber\\
B_{iiii} &=&  a_i^4 B_{0000}, \label{Abif}\\
B_{ijij} &=& 2 C_{0i0i} C_{0j0j} \label{Abij}\\
&=& \frac{1}{18 a_k^2}\Bigl(2{\dot a_i}a_j {\dot a_k} -a_i {\dot
a_j} {\dot a_k} - {\dot a_i}{\dot a_j} a_k - a_i a_j {\ddot a_k} -
{\ddot a_i} a_j a_k + 2 a_i {\ddot a_j} a_k\Bigr)\nonumber \\
&\times& \Bigl(2 a_i {\dot a_j} {\dot a_k} - {\dot a_i} a_j {\dot
a_k} - {\dot a_i}{\dot a_j} a_k - a_i a_j {\ddot a_k} - a_i {\ddot
a_j} a_k + 2 {\ddot a_i} a_j a_k\Bigr), \nonumber\\ B_{0k0k} &=& -
\frac{a_k^2}{a_i^2 a_j^2} B_{ijij}. \quad i,j,k = 1,2,3, \quad i
\ne j \ne k. \label{Ab0k}
\end{eqnarray}
\end{subequations}

Thus we have used three different definition of BR. The first one
defined in \cite{Deser} imposes some restriction on the metric
functions, namely for the BI spacetime is coincides with vacuum
solution of Einstein equations. The second definition removes this
restriction, but since BI metric admits non-trivial Ricci tensor,
the BR in this case is not totally symmetric. Finally we gave the
definition used by Bergqvist and Senovilla. It satisfies all the
properties of BR and it is totally symmetric. In what follows we
study the DEP and DSEP for BR in BI Universe.

\section{Cosmological singularity and the dominant energy
condition}

Recalling that a timelike geodesic is a world line for a particle
moving without acceleration in the proper reference system we
define the following:

A spacetime is nonsingular if any timelike geodesics, or null
geodesics, can be continued into the past and the future without
bound, i.e., to infinite proper length for the timelike geodesics
and to an infinite value of an affine parameter for the null
geodesics. Such a spacetime is termed "causally, geodesically
complete". The requirements on the completeness are the minimum
necessary so that the spacetime does not contain a singularity. It
is necessary to point out that a spacetime not satisfying these
requirements, however, one with a singularity, does not
necessarily contain points with infinite curvature or with small
hole.

{}From physical point of view, of course, one ought to take as
singular any spacetime in which the geodesic world line of a
particle cannot be continued without bound with respect to the
proper time of this particle, for such a singular spacetime would
lead to a violation of conservation laws.

As applied to the cosmological problem, the Hawking-Penrose theorem
reads as follows~\cite{hawking}:

{\bf Theorem.}
{\it A space-time} ${\mathcal M}$
{\it cannot satisfy causal
geodesic completeness if the GTR (General Theory of Relativity) equations 
hold and if the following conditions are fulfilled:

(i) The space-time} ${\cal M}$ {\it does not contain closed 
time-like curves.

(ii) The energy condition (DEP) is satisfied
at every point.} \\
The energy condition may be expressed as
\begin{equation}
t^\alpha t_\alpha = 1 \quad {\rm implies} \quad R_{\alpha \beta}
t^\alpha t^\beta \le 0. \label{dec00}
\end{equation}
With Einstein's equations
\begin{equation}
R_{\alpha\beta} - \frac{1}{2} g_{\alpha \beta} R = - \kappa
T_{\alpha \beta}, 
\nonumber
\end{equation}
(\ref{dec00}) becomes
\begin{equation}
t^\alpha t_\alpha = 1 \quad {\rm implies} \quad T_{\alpha \beta}
t^\alpha t^\beta \ge \frac{1}{2} T^\mu_\mu. \label{dec0}
\end{equation}
If, in an eigentetrad  of $T_{\mu \nu}$, $\ve$ denotes the energy
density and $p_1,\,p_2,\,p_3$ denote the three principal pressure,
then (\ref{dec0}) can be written as
\begin{eqnarray}
\ve + \sum_{\alpha} p_\alpha &\ge& 0;\no 
\ve + p_\alpha &\ge& 0,
\quad \alpha = 1,2,3. \nonumber
\end{eqnarray}
The weak energy condition is
\begin{equation}
\ell^\alpha \ell_\alpha = 0 \quad {\rm implies} \quad R_{\alpha
\beta} t^\alpha t^\beta \le 0, 
\nonumber
\end{equation}
which is a consequence of (\ref{dec00}).

{\it (iii) On each time-like or null geodesic $\gamma$, there
is at least one point for which}
\begin{equation}
K_{[a}R_{b]cd[e}K_{f]}K^c K^d \ne 0, \label{c3} 
\end{equation}
where $K_a$ is the tangent to the curve $\gamma$ at the given
point and where the brackets on the subscripts imply
antisymmetrization. If $\gamma$ is timelike, we can rewrite
(\ref{c3}) as
\begin{equation}
R_{abcd} K^c K^d \ne 0. 
\nonumber
\end{equation}

{\it (iv) The space-time} ${\cal M}$ {\it contains either $(a)$ a trapped
surface, $(b)$ a point $P$ for which the convergence of all the
null geodesics through $P$ changes sign somewhere to the past of
$P$, or $(c)$ a compact space-like hypersurface.}

The DEP for the BI metric can be written in
the form:
\begin{subequations}
\begin{eqnarray}
T_{0}^{0} &\ge& T_{1}^{1} a^2 + T_{2}^{2} b^2 + T_{3}^{3} c^2,\\
T_{0}^{0} &\ge& T_{1}^{1} a^2, \\ T_{0}^{0} &\ge& T_{2}^{2} b^2,\\
T_{0}^{0} &\ge& T_{3}^{3} c^2.
\end{eqnarray}
\end{subequations}

In analogy with the Hawking-Penrose theorem stated above,
Senovilla and others introduced DEP for the higher
dimensional tensors. A detailed description of singularity
theorems and their consequences can be found in the review paper
by Senovilla \cite{sengrg}.

The DSEP as defined by Senovilla reads:
\cite{Sen1}

{\bf Theorem.} {\it A rank-$s$ tensor} $T_{\mu_1...\mu_s}$,
{\it is said to satisfy the DSEP if}
\begin{equation}
T_{\mu_1...\mu_s}k_1^{\mu_1}...k_s^{\mu_s} \ge 0 \label{DPdef}
\end{equation}
{\it for any future-pointing causal vectors}
$k_1^{\mu_1}...k_s^{\mu_s}$.\\ To justify its name the dominant
DSEP obeys the following Lemma.

{\bf Lemma.} {\it If a tensor} $T_{\mu_1...\mu_s}$
{\it satisfies the {\rm DSEP}, then}
\begin{equation}
T_{0...0}\ge |T_{\mu_1...\mu_s}|, \quad \forall \mu_1,...,\mu_s =
0,...,n-1 \label{DPLem}
\end{equation}
{\it in any orthonormal basis} $\{\vec{e_\nu}\}$.

It was also established in \cite{bergsen}, that any tensor
satisfying the DEP possesses the following property:

{\bf Property.} $T_{\mu_1...\mu_s}$ {\it satisfies DEP if and 
only if}
\begin{equation}
T_{\mu_1...\mu_s} l_1^{\mu_1}...l_s^{\mu_s} \ge 0 \label{nulvec}
\end{equation}
{\it for any set} $l_1^{\mu_1}...l_s^{\mu_s}$ {\it of
future-pointing null vectors.}

Let us now back to the 4-rank BR tensor and to check the DSEP for 
it. Since BR defined as
(\ref{BelW1}) is a completely symmetric, trace-free 4-rank tensor,
then it satisfy the DEP \cite{BLan}. Therefore from the foregoing
theorem, lemma and property for $B_{ijkl}$ we can write:

\begin{equation}
B_{abcd}k_1^a k_2^b k_3^c k_4^{d} \ge 0, 
\nonumber
\end{equation}
\begin{equation}
B_{0...0}\ge |B_{abcd}|, \quad \forall a,b,c,d = 0,1,2,3
\label{DPLembr}
\end{equation}
and
\begin{equation}
B_{abcd} l_1^a l_2^b l_3^c l_4^d \ge 0 
\nonumber
\end{equation}

Apparently, in view of (\ref{Abif}), Eq. (\ref{DPLembr}) imposes 
some restriction on the metric functions, e.g., $a_i^4 \le 1$.
But this is not the case, since we have used a coordinate basis 
to compute the BR tensor components (\ref{Ab0f0})-(\ref{Ab0k})
and Eq. (\ref{DPLem}) refers to an orthonormal basis. We
recall that in an orthonormal basis the components of the Weyl 
tensor obey the following relation:
\begin{equation}
C_{0i0i} = - C_{jkjk}, \quad i,j,k = 1,2,3 
~{\rm and}~ i \ne j \ne
k. \label{weylr}
\end{equation}
In view of \eqref{weylr} the expressions \eqref{Ab0f0},
\eqref{Abif}, \eqref{Abij} and \eqref{Ab0k} now read
\begin{subequations}
\begin{eqnarray}
B_{0000} &=& \Bigl(C_{jkjk}\Bigr)^2 + \Bigl(C_{kiki}\Bigr)^2 +
\Bigl(C_{ijij}\Bigr)^2 \label{Ab0f0on} \\
B_{iiii} &=&  B_{0000}, \label{Abifon}\\
B_{ijij} &=& 2 C_{0i0i} C_{0j0j} \label{Abijon}\\
B_{0k0k} &=& - B_{ijij}. \quad i,j,k = 1,2,3, \quad i \ne j \ne k.
\label{Ab0kon}
\end{eqnarray}
\end{subequations}

Thus in connection with the above Lemma relative to DSEP, Eq. 
(\ref{DPLembr}) is fulfilled without any restrictions on the 
metric functions.

\section{Conclusions}

In view of the importance of the BI model in the study of the
present day Universe we considered the most simple model with a
perfect fluid as a source field. The corresponding solutions to
the Einstein equations have been obtained. Three alternative
definitions of Bel-Robinson tensor are considered. It is shown
that the definition used by Deser {\it et. al.} is consistent with
the Einstein equations when the source field is given by a vacuum
only. The second definition used by Teyssandier is free from this
restriction, but BR defined in this way is not totally symmetric.
Definition used by Senovilla and Bergqvist does not suffer from
this shortcomings, i.e., it has all the symmetries and the DSEP is 
 satisfied.

\subsection*{Acknowledgments}
Part of the present work was done with the financial support from
a grant afforded by the Representative of Romanian Government to
JINR-Dubna within the Hulubei-Meshcheryakov Programme on 2006. M.
V. has been partially supported by a grant CNCSIS, 
Romania. We would also like to thank J. M. M. Senovilla for
valuable remarks.
\newpage
\end{document}